\documentclass[reprint, superscriptaddress, amsmath, amssymb, aps, nofootinbib]{revtex4-2}
\usepackage{float}
\usepackage{graphicx}
\usepackage{flushend}
\usepackage{placeins}
\usepackage{dcolumn}
\usepackage{bm}
\usepackage[colorlinks=true]{hyperref}
\usepackage{booktabs}
\usepackage{afterpage}
\usepackage{balance}
\usepackage[normalem]{ulem}

\begin{document}

\preprint{APS/123-QED}

\title{New modeling for hybrid stars with an elastic quark core}

\author{Zoey Zhiyuan Dong}
\email{dongzhiyuan@mails.ccnu.edu.cn}
\affiliation{Institute of Astrophysics, Central China Normal University, Wuhan 430079, China}

\author{Shu Yan Lau}
\email{shuyan.lau@montana.edu}
\affiliation{Department of Physics, Montana State University, Bozeman, MT 59717, USA}
\affiliation{Department of Physics, University of Virginia, Charlottesville, Virginia 22904, USA}

\author{Kent Yagi}
\email{ky5t@virginia.edu}
\affiliation{Department of Physics, University of Virginia, Charlottesville, Virginia 22904, USA}

\date{\today}

\begin{abstract}
Heavy neutron stars may contain solid quark cores as motivated by, e.g., the crystalline color superconducting phase, forming \emph{elastic hybrid stars} (HSs). Many previous studies assumed an elastic core to be unsheared for the background, static and spherically symmetric configuration, and introduced shear deformation only at a perturbative level. This study relaxes this assumption and explores the influence of nonlinear elasticity on the static, spherically symmetric structure of elastic HSs within a fully relativistic elasticity framework. Such a framework effectively introduces anisotropic pressure within the quark matter core due to elasticity. The quark core is modeled using a quasi-Hookean equation of state (EOS) with shear contributions, while the nuclear matter envelope is treated as a perfect fluid. We find that including elasticity increases the maximum mass of HSs by several percent. This enhancement allows some soft EOSs to satisfy current observational constraints. However, since the effects of elasticity are primarily concentrated in the high-mass regime, the current observational constraints are insufficient to distinguish whether an elastic anisotropic quark core exists within these stars. Additionally, we show that the compactness of stable stars can exceed the critical value of 1/3 due to the inclusion of elasticity, making them potential candidates for black hole mimickers. Furthermore, we found that common phenomenological anisotropy models fail to accurately describe the anisotropy of the elastic core and propose a new parametrized anisotropy model that can accurately capture physically-motivated profiles with an error of 10\% across a wide parameter space. This work not only bridges the gap between elastic EOSs and parametrized anisotropic models but also provides a foundation for interesting applications such as studying nonradial perturbations, tidal deformability, and pulsation modes for elastic HSs.
\end{abstract}

\maketitle
\section{\label{Sec: intro}Introduction}
The internal structure of neutron stars (NSs) remains an open question in astrophysics. One of the candidates for the inner core composition is quark matter (QM), which is predicted to exist at extreme densities where the quarks become asymptotically free. If the core consists of QM while the outer layers remain composed of hadronic matter and there is a phase transition in between, the object is referred to as a hybrid star (HS) to distinguish it from the typical NSs with a hadronic core. 

Although typical NSs and HSs have been modeled as isotropic perfect fluids, they may contain stress anisotropy of various astrophysical origins. One such possibility is the existence of elastic layers within the stars. An NS crust can have a shear modulus of $10^{29}~\text{erg cm}^{-3}$ \cite{Douchin_2000, Haensel_2007}, and is crucial to explain pulsar glitches. Since this value is much smaller than the bulk modulus, it is not expected to have a significant impact on the NS structure. 
However, it can still affect the perturbed properties like the static tides \cite{Gittins_2020, Pereira_2020} and the oscillation modes \cite{McDermott_1988, Kruger_2015}. The existence of the elastic crust can further lead to possible electromagnetic observables due to crustal failure \cite{Tsang_2012}. This potentially allows one to probe the existence of a sharp phase transition in the NS interior \cite{Pereira_2023}. Besides, the crust can support mountains which source gravitational waves from spinning NSs \cite{Haskell_2006, Ushomirsky_2000}.

Meanwhile, the QM component of an HS can be in a crystalline solid state, such as the crystalline color superconducting (CCS) phase, due to the formation of Cooper pairs with non-vanishing momenta via a non-BCS mechanism \cite{Alford_2001, Alford_2001_2, Rajagopal_2002} or the formation of quark clusters \cite{Xu_2003}. The shear modulus of the QM in the HS core is predicted to be on the order $10^{32}~\text{erg cm}^{-3}$ or higher. 
In addition, the hadron–quark mixed phase may itself crystallize with an estimated shear modulus of about a few times $10^{33}~\text{erg cm}^{-3}$\cite{Johnson-McDaniel_2012} and thus can be viewed as another potential elastic layer.
Such high shear moduli can cause the structure of HSs to differ substantially from that of an isotropic fluid star, even under spherical symmetry configuration, when nonzero shear strain is considered.
Moreover, the large shear modulus can also influence HS astrophysical properties, such as quadrupolar deformations\cite{Haskell_2007, Johnson-McDaniel_2013, Lau_2017, Lau_2019}, rotational frequency \cite{Anglani_2014}, or oscillation modes \cite{Lin_2013, Lau_2018}. To investigate these properties, an elastic theory compatible with relativity is necessary.

An early reference for relativistic elasticity is given by Carter and Quintana \cite{Carter:1972}. In particular, their formalism covers elasticity in the finite-strain regime applicable to the high-pressure environment of the compact star interior. While there are various studies of self-gravitating elastic spheres in the context of relativity \cite{Karlovini:2002fc, Brito_2010, Brito_2011, Natrio_2020, Alho_2022}, they are rarely linked to the possible elastic phases within an HS.
Interestingly, Alho \textit{et al.} \cite{Alho_2022} demonstrated, by introducing a polytropic model extended to include elasticity, that some elastic stars with a radius within the (unstable) light ring can be dynamically stable against radial perturbations and thus are potential black hole mimickers. However, they might still be subject to nonlinear instabilities \cite{Cardoso_2014, Keir_2016}.
On the other hand, many other astrophysically relevant studies on the elastic properties of NSs or HSs usually assume that the spherically symmetric structure is unsheared and elasticity enters only when the star is non-radially deformed \cite{McDermott_1988, Finn_1990, Yoshida_2002, Haskell_2007, Penner_2011, Kruger_2015, Pereira_2019, Pereira_2020, Andersson_2021, Pereira_2021, Pereira_2023}. This assumption does not hold, in general, and can lead to a significant offset in the background structure of the star if the shear modulus is large.

Elasticity introduces modifications to the properties of NSs and HSs, particularly in systems with crystalline QM cores. The core’s resistance to shear deformation, quantified by its shear modulus $\tilde{\mu}$, ranges from 20 to 1000 times greater than that of the nuclear matter crust~\cite{Mannarelli_2007}. This enhanced rigidity affects the mass, radius, and tidal deformability of HSs compared to their fluid counterparts \cite{Dong2024, Pereira_2021}.
Such effects become particularly relevant in analyzing observations like GW170817~\cite{PhysRevLett.120.172703, PhysRevLett.119.161101, Abbott_2018, PhysRevX.9.011001}, which provide direct constraints on tidal deformability, and the electromagnetic measurements of mass-radius ($M$--$R$) relations~\cite{Miller_2019, Miller_2021, Salmi_2024, Choudhury_2024, Vinciguerra_2024}. For a detailed overview of recent progress, see e.g.~\cite{chatziioannou2024review}.

Instead of focusing on the anisotropic stress from elasticity, anisotropic compact stars are also constructed using parametrized models for the pressure anisotropy as a function of the energy density. This approach, similar to the equation of state (EOS) relating fluid pressure and energy density in isotropic perfect-fluid stars, was first taken by Bowers and Liang \cite{Bowers_Liang1974} and was followed by various studies \cite{Cosenza1981, Bayin_1982, Hillebrandt_1976, Dev_2003, Horvat_2011, Arbail_2016, Isayev_2017, Pretel2020}.
These models assume dependence not only on the thermodynamic state variables but also explicitly on other quantities, such as the radial coordinate $r$ or the mass $m(r)$ enclosed within a sphere of radius ``$r$''. Such dependence is necessary to make the anisotropy vanish at the stellar center and avoid a singularity~\cite{Bowers_Liang1974}. Since they involve dependence other than the state variables, they cannot be considered as EOSs in the strict sense.
Furthermore, these models are somewhat unsatisfactory in that the ``cure'' to the singularity is enforced by hand without fundamental connections to matter properties. They also do not inform us of the change in anisotropy when the star is perturbed from the spherically symmetric configuration.
Moreover, the connections between these effective models and the underlying origin of anisotropy are unclear \cite{Poisson_2024}.

The main goal of this study to advance our understanding of anisotropic effects in HSs is two-fold. The first one is to relax the assumption in previous literature and include the effect of elasticity within the spherically symmetric configuration to see how it affects global stellar quantities, such as the mass and radius. The second one is to establish connections between the parametrized model of anisotropy and realistic elastic EOSs. We construct static spherically symmetric HS models with an elastic QM core from the Einstein field equations, assuming the solid QM follows a specific form of \emph{quasi-Hookean} EOS first introduced in \cite{Karlovini:2002fc}. The models are thus consistent with both general relativity and a specific class of elastic material properties\footnote{Cadogan and Poisson~\cite{Poisson_2024, Cadogan:2024ohj, Cadogan:2024ywc} recently established a different, relativistic framework for self-gravitating anisotropic fluids based on a theory of liquid crystals.}. We call these types of HSs \emph{elastic} HSs, while those composed entirely of isotropic fluid are \emph{fluid} HSs. Using these results, we introduce a parametrized model that can approximate the pressure anisotropy in the elastic QM core well. This provides a direct relation between the free model parameters and the physical quantities of an elastic solid.

Our main findings can be summarized as follows:
\begin{enumerate}
\item Figure~\ref{fig: MR} presents the mass-radius relation for elastic HSs for various shear modulus coefficients $\kappa$ that is defined as the ratio between the shear modulus $\tilde \mu$ and the square root of the energy density. Observe that the presence of an elastic core can increase the maximum mass of HSs. However, it remains insufficient to distinguish whether HSs possess an elastic core from LIGO/Virgo and NICER/XMM-Newton.

\item
Figure~\ref{fig: sigma compares} compares the pressure anisotropy profile for the core of an elastic HS, obtained numerically, against various phenomenological anisotropy models. Observe that common models like the ones proposed by Bowers and Liang (BL model)~\cite{Bowers_Liang1974} and Horvat \emph{et al.} (H model)~\cite{Horvat_2011} fail to accurately capture the numerical data from a physically-motivated elastic HS configuration (see Appendix~\ref{AP: fitting example} for more details on the comparison). We thus provide the following new phenomenological model to more accurately describe the realistic profile: 
\begin{align}
    \sigma_{\mathrm{fit}}=\lambda p_c\left(\frac{\rho}{\rho_c}\right)^{\frac{N}{\lambda}-1} \frac{m}{r}\left(1-\frac{2 m}{r}\right)^{-1}, \label{eq: sigma fit}
\end{align}
where $\rho$ is the energy density while $\lambda$ and $N$ are dimensionless values that can be expressed in terms of the shear modulus coefficient $\kappa$ and the central pressure $p_c$. This fitting achieves an accuracy with an error of 10\%, as shown in the bottom panel of Fig.~\ref{fig: sigma compares}.

\begin{figure}[!htp]
    \centering
    \includegraphics[width=8.6cm]{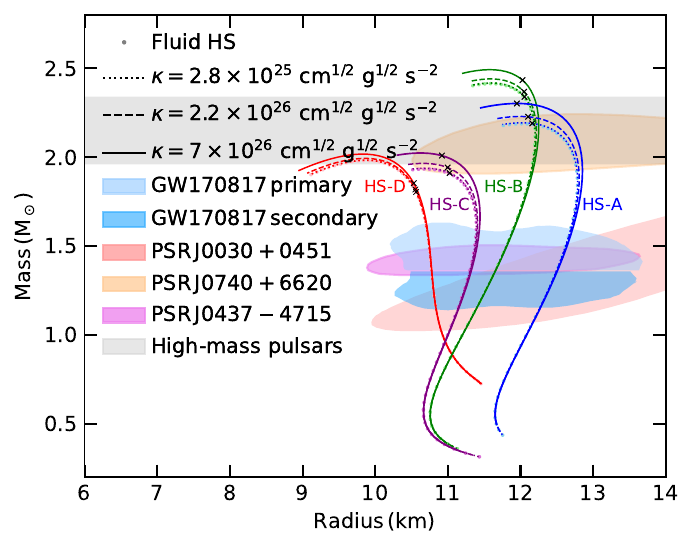}
    \caption{\label{fig: MR}
    Mass-radius relations for HS models constructed with EOS HS-A (blue), HS-B (green), HS-C (purple), and HS-D (red), whose base nuclear matter EOS is the APR EOS. The detailed EOS parameters can be found in Table~\ref{Table: HS EOSs}. The shear modulus coefficient $\kappa$ varies among $2.8 \times 10^{25}\text{cm}^{1/2}\text{g}^{1/2}\text{s}^{-2}$ (dotted), $2.2 \times 10^{26}\text{cm}^{1/2}\text{g}^{1/2}\text{s}^{-2}$ (dashed), and $7 \times 10^{26}\text{cm}^{1/2}\text{g}^{1/2}\text{s}^{-2}$ (solid). The isotropic fluid HSs are represented by light-color dots. The cross symbols mark stars with $p_c = 4.23 \times 10^{35}\text{dyn cm}^{-2}$ in the stable branches. 
    Various observational bounds are shown with shaded regions, including the mass and radius measurement of GW170817 \cite{Abbott_2018} (90\% confidence interval, light blue and blue regions), J0030+0451 with NICER (95\% confidence interval, maroon region) \cite{Vinciguerra_2024}, J0740+6620 with NICER and XMM-Newton (95\% confidence interval, orange region) \cite{Salmi_2024}, J0437-4715 with NICER (95\% confidence interval, violet region) \cite{Choudhury_2024}, and the mass measurement of J0740+6620 from pulsar timing \cite{Cromartie_2020} (95.4\% confidence interval, grey horizontal band).}
\end{figure}

\begin{figure}[!htp]
    \centering
    \includegraphics[width=8.6cm]{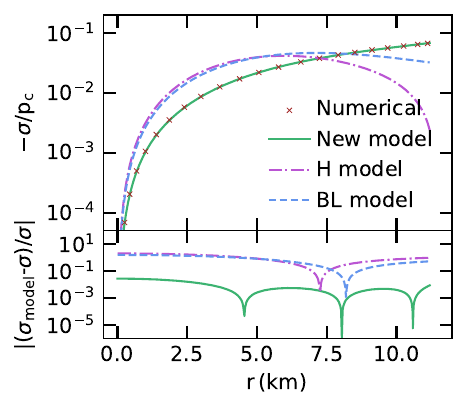}
    \caption{\label{fig: sigma compares} (Top) The $\sigma$ profile of the HS-A model (brown crosses) is fitted by our model (green solid), H model (purple dot-dashed), and BL model (blue dashed). We fix ($p_c$, $\kappa$) = ($4.23 \times 10^{35} \mathrm{~dyn} \mathrm{~cm}^{-2}$, $7 \times 10^{26} \, \mathrm{cm}^{1/2} \, \mathrm{g}^{1/2} \, \mathrm{s}^{-2}$). The fitted anisotropy parameters for the H and BL models are given by $\lambda_\mathrm{H} = -0.315108$ and $\lambda_{\mathrm{BL}} = -0.803287$, respectively. (Bottom) Fractional difference between analytical models and the numerical results.}
\end{figure}
\end{enumerate}

The remainder of this paper is structured as follows. In Section~\ref{Sec: Formulation}, we review the formulation to construct elastic HSs derived by \cite{Karlovini:2002fc}. Section~\ref{Sec: Stellar Structure} is dedicated to exploring the effect of elasticity on the stellar structure. Section~\ref{Sec: Anisotropic Profile Modelling} introduces a new parametrized model to capture the nature of anisotropy in the elastic core. Finally, in Section~\ref{Sec: Conclusion}, we summarize our findings and discuss the broader implications of our results for understanding anisotropic HSs. Unless otherwise stated, we adopt the geometric units $c = G = 1$ throughout the paper.

\section{\label{Sec: Formulation}Formulation of anisotropic hybrid stars}
The elastic core of the HS is under high pressure and, in general, has a non-negligible shear deformation from the relaxed state. In many previous studies, the solid component of a static spherically symmetric star is assumed to be unsheared, and a linear strain-stress relation (i.e., the Hookean relation \cite{Carter:1972, Andersson_2021}) is assumed to hold when the star is perturbed (e.g., \cite{McDermott_1988, Finn_1990, Yoshida_2002, Penner_2011, Kruger_2015, Lau_2017, Lau_2019, Gittins_2020, Dong2024}). 
The validity of this assumption relies heavily on the small deviation of the background configuration from the true unsheared state of the solid, which, in general, does not hold. As a result, this treatment has ignored the background shear that can impact the stellar structure. Here, we focus on the effect of elasticity on the static structure of HSs by employing a fully relativistic nonlinear elasticity theory developed in \cite{Carter:1972}. 
We further assume the elastic core can be described by the quasi-Hookean EOS studied in \cite{Carter:1972, Karlovini:2002fc}, which we shall give a brief review in this section. 

\subsection{\label{Sec: EOS}Equation of State}
In terms of notation, we use an overhead tilde to represent the unsheared components, depending only on the particle number density $n$. Relations among these unsheared quantities correspond to what is typically referred to as the fluid EOS, which depends only on volume changes. 

For elastic matter, the EOS should also account for shear deformation. We use the quasi-Hookean EOS, which relates to a shear scalar, describing how the energy density depends on the magnitude of shear deformation \cite{Karlovini:2002fc}
\begin{align}
    \rho&=\tilde{\rho}+\tilde{\mu} \mathcal{S}^2\,,
    \label{eq: quasi-Hookean EOS}
\end{align}
where $\rho$ is the energy density, $\tilde{\mu}$ is the shear modulus, and $\mathcal{S}^2$ is the shear scalar, obtained from the contractions of the shear tensor\footnote{The contraction can be quadratic between the shear strain tensor itself, e.g., by $s_{ab}s^{ab}$, or with the tensor representing the unsheared state (see Eq. (5.1) of \cite{Carter:1972} for definition), or even a cubic contraction as the one we employ here \cite{Karlovini:2002fc}.}, $s_{ab}$, that quantifies the shear deformation (see, e.g., Eq. (5.17) of \cite{Carter:1972} for its definition\footnote{In linear elasticity, the shear strain tensor is simply the traceless part of the infinitesimal strain tensor. In nonlinear cases, the shear strain is no longer traceless.}).
There are many proposed choices of $\mathcal{S}^2$, and its actual form depends on the material (see~\cite{Alho_2024} for more examples). Here, we adopt the form introduced in \cite{Karlovini:2002fc} as explained in the following paragraph. 

The shear scalar $\mathcal{S}^2$ of a quasi-Hookean material measures the deformation that goes beyond simple volume changes. It remains zero for pure volume changes from the relaxed state but takes nonzero values when shear strain is introduced. We follow the formalism derived in~\cite{Karlovini:2002fc}, where the shear scalar $\mathcal{S}^2$ is defined as
\begin{align}
    \mathcal{S}^2=\frac{1}{12}\left[\left(\frac{n_1}{n_2}-\frac{n_2}{n_1}\right)^2+\left(\frac{n_2}{n_3}-\frac{n_3}{n_2}\right)^2+\left(\frac{n_3}{n_1}-\frac{n_1}{n_3}\right)^2\right]. \label{eq: shear scalar}
\end{align}
Here, $n_1$, $n_2$, and $n_3$ are the linear particle densities in the three orthogonal principle directions, i.e., the number of particles contained in a small unit volume divided by the unit length in that direction (see \cite{Karlovini:2002fc} for the precise definition), which represent how the material is stretched or compressed along their principal axes when compared to the relaxed state. If $n_1=n_2=n_3$, the material is unsheared, meaning it deforms equally in all directions relative to the relaxed state. Specifically, the shear scalar should meet criteria that include consistency with the quadratic invariant of the shear tensor in the linear elasticity regime, as well as ensuring that $\mathcal{S}^2$ remains a non-negative invariant under coordinate transformations. 
As a result, we can recover the linear Hookean relation (the linear stress-strain relation or Hooke's law) from Eq.~\eqref{eq: quasi-Hookean EOS} for small deformations about the unsheared state.
The definition in Eq.~\eqref{eq: shear scalar} is somewhat simpler than the one used by Carter and Quintana~\cite{Carter:1972} in the sense that it does not contain fractional power in the linear number densities.

Let us further rewrite the above shear scalar $\mathcal{S}^2$.
We introduce the linear particle densities in the radial direction $n_r$ and the tangential direction $n_t$ to describe spherically symmetric solids. In such cases, we can substitute $n_1 = n_r$ and $n_2=n_3=n_t$ in Eq.~\eqref{eq: shear scalar} since we assume the deformation along the tangential directions is degenerate. Here, following \cite{Karlovini:2002fc}, we define $ z = n_r / n_t$, and the shear scalar simplifies to
\begin{align}
    \mathcal{S}^2=\frac{1}{6}\left(z^{-1}-z\right)^2\,.\label{eq: shear scalar in z}
\end{align}

\begin{figure}[!htp]
    \centering
    \includegraphics[width=8.6cm]{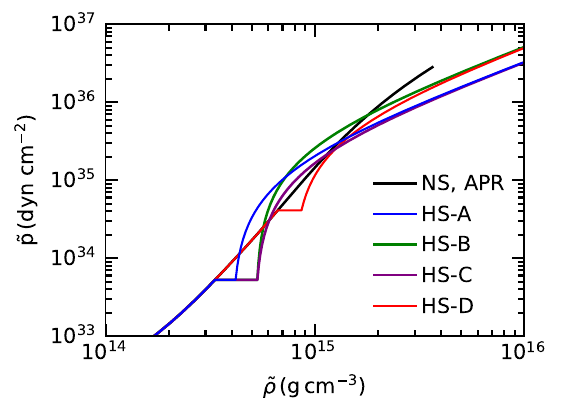}
    \caption{\label{fig: EOS} 
    The EOSs for NSs and HSs with various EOS parameters. The APR EOS, shown by the black line, serves as a reference to NM EOS. The EOS parameters for each HS model are listed in Table~\ref{Table: HS EOSs}. We will use the legends introduced here (i.e., HS-A) to denote the HS models used in our analysis. 
    }
\end{figure}

In our study, the fluid EOS, the relation between unsheared pressure $\tilde{p}$ and unsheared energy density $\tilde{\rho}$, incorporates both the nuclear matter (NM) and QM regions, connected by a sharp phase transition at a transition pressure $p_\mathrm{trans}$. We use the tabulated APR \cite{PhysRevC.58.1804} to describe the NM component, denoted as $\tilde{\rho}_{\mathrm{NM}}(\tilde{p})$. The QM core is described using the Constant Speed of Sound (CSS) template. The resulting EOS is given by \cite{PhysRevD.88.083013}
\begin{align}
    \tilde{\rho}(\tilde{p})= \begin{cases}\frac{\tilde{p}-p_\mathrm{trans}}{\tilde{v}^2}+\tilde{\rho}_{\mathrm{NM}}\left(p_\mathrm{trans}\right)+\Delta \tilde{\rho}\,,&\text {if } p_r>p_\mathrm{trans} \\ \tilde{\rho}_{\mathrm{NM}}(\tilde{p})\,, &\text {if } p_r \leq p_\mathrm{trans}\end{cases}\label{eq: EOS}
\end{align}
where the fluid sound speed $\tilde{v}$ is assumed to be a constant while $\Delta \tilde{\rho}$ is the energy density gap between the QM and NM. Note that the condition distinguishing these two regions is defined in terms of the radial pressure $p_r = n n_r \partial (\rho/n)/\partial n_r$ because the core region is anisotropic while $p_r$ reduces to $\tilde{p}$ in the fluid envelope region. Since $p_\mathrm{trans}$ represents a specific value at the boundary of the EOS, we do not distinguish it with or without a tilde notation. We will discuss the boundary conditions, including the one at the QM-NM interface, in Sec.~\ref{Sec: BC}. Figure~\ref{fig: EOS} provides examples of the HS model for different values of $p_\mathrm{trans}$, $\Delta \tilde{\rho}$, and $\tilde{v}$. The unsheared energy density for NM in APR EOS exhibits a log-linear relation with unsheared pressure. 

\begin{table}[h]
    \centering
    \begin{tabular}{cccc}
        \toprule
         & $p_\mathrm{trans}~(\text{dyn} \text{~cm}^{-2})$ & $\tilde{v}^2$ &  $\Delta \tilde{\rho}~(\text{g}\text{~cm}^{-3})$ \\ 
        \midrule
        HS-A & $5.26 \times 10^{33}  $ & 0.38 & $8.50 \times 10^{13}  $ \\  
        HS-B & $5.26 \times 10^{33}  $ & 0.60 & $1.98 \times 10^{14}  $ \\ 
	HS-C & $5.26 \times 10^{33}  $ & 0.38 & $1.98 \times 10^{14}  $\\ 
        HS-D & $4.11 \times 10^{34}  $ & 0.60 & $1.98 \times 10^{14}  $ \\
        \bottomrule
    \end{tabular}
    \caption{HS EOSs parameters studied in this paper. These models satisfy all the bounds on the HS EOSs from LIGO/Virgo, NICER, and Chandra Collaborations mentioned in Sec.~\ref{Sec: EOS}, as well as the mass-radius constraints in Fig.~\ref{fig: MR}, that were obtained by assuming non-elastic, fluid models.}
    \label{Table: HS EOSs}
\end{table}

The CSS serves as a parameterization template and was developed to avoid model dependence and to offer flexibility in describing various configurations of NSs (the well-known MIT Bag model \cite{Chodos_1974} is a good example of this template).
The parameters of CSS templates, such as $\Delta \tilde{\rho}$, $p_\mathrm{trans}$, and $\tilde{v}$, are designed as convenient tools for describing stellar properties rather than representing any direct physical quantities. It allows for flexibility in modeling stars where $\tilde{v}$ changes gently, making it versatile for different types of compact stars, including those undergoing first-order phase transitions. However, since all wave speeds in the solid must satisfy the causality limit, there are further restrictions on the fluid sound speed. We choose 0.8 as an upper bound for $\tilde{v}^2$, and the details will be discussed in Appendix~\ref{Sec: Wave speed}. 

In addition to the causality limit, there are other constraints on the EOS. For instance, based on observations from GW170817, which involved a binary neutron star system, an analysis was conducted using a parameterized EOS \cite{PhysRevLett.121.161101}. This analysis made use of the EOS-insensitive relation between various properties of NSs \cite{YAGI20171, Yagi_2016, Chatziioannou_2018, Maselli_2013, Urbanec_2013}. It was found that, at twice nuclear saturation density, the pressure is constrained at the 90\% confidence level to  $3.5^{+2.7}_{-1.7} \times 10^{34} \text{~dyn} \text{~cm}^{-2}$.   
Moreover, in \cite{PhysRevC.103.035802}, the authors used the CSS template to analyze the parameter space of EOS based on data from LIGO/Virgo \cite{PhysRevLett.121.161101, De2018}, NICER \cite{Riley_2019, Miller_2019}, and Chandra Collaborations \cite{Lattimer2014}. They emphasized that under the Seidov stability condition \cite{Seidov_1971}, the EOS parameters satisfy $\rho_t/\rho_0 = 1.6^{+1.2}_{-0.4}$ and $\Delta \epsilon/\epsilon_t = 0.4^{+0.20}_{-0.15}$ at 68\% confidence level. All of our EOS models are viable candidates that simultaneously satisfy the EOS constraints discussed above and the observational constraints on the mass and radius presented in Fig.~\ref{fig: MR}.

A key property distinguishing elastic HSs from fluid HSs is the shear modulus of the QM phase in the core. The shear modulus quantifies the rigidity of the elastic core, allowing it to sustain anisotropic stress and thereby influencing the star’s overall structure and observable properties. To model this effect, we adopt a shear modulus $\tilde{\mu}$ that depends on the energy density as follows:
\begin{align}
    \tilde{\mu}=\kappa \sqrt{\tilde{\rho}}\,,\label{eq: shear modulus}
\end{align}
with a constant $\kappa$. Such a simple form is inspired by the calculation of the shear modulus of the CCS phase derived in \cite{Mannarelli_2007}:
\begin{align}\label{eq: ccs mu}
    \tilde{\mu}_{\mathrm{CCS}}=2.47 \mathrm{MeV}\,\mathrm{fm}^{-3}\left(\frac{\Delta}{10 \mathrm{MeV}}\right)^2\left(\frac{\mu_q}{400 \mathrm{MeV}}\right)^2,
\end{align}
where $\Delta$ is the gap parameter and $\mu_q$ represents the chemical potential of QM. 
In the high-density limit, $\tilde{\mu}_\text{CCS} \propto \mu_q^2$, and $\tilde{\rho} \propto \mu_q^4$ as for the ultrarelativistic free Fermi gas. Hence, we have $ \tilde{\mu}_\text{CCS} \propto \sqrt{\tilde{\rho}}$ and $\kappa \propto \Delta^2$. Mannarelli \textit{et al}.~\cite{Mannarelli_2007} placed constraints on the gap parameter $\Delta$ in the CCS phase, suggesting a range of approximately 5 MeV to 25 MeV. Therefore, in our approximation, $\kappa$ is set to lie within the range\footnote{See Eqs.~(12)--(16) in \cite{Faggert:2023kvm} for the derivation of the range for $\kappa$.} $2.8 \times 10^{25} \, \text{cm}^{1/2} \, \text{g}^{1/2} \, \text{s}^{-2}$ to $7 \times 10^{26} \, \text{cm}^{1/2} \, \text{g}^{1/2} \, \text{s}^{-2}$, ensuring that the quark core’s rigidity is accounted for in line with the current theoretical understanding. While these constraints on $\kappa$ ensure consistency with current theoretical understanding, we note that Eq.~\eqref{eq: shear modulus} can still lead to an overestimation compared to Eq.~\eqref{eq: ccs mu}, especially in the low-density regime.

\subsection{\label{Sec: Structural Eq}Structural Equations}
We now turn our attention to the sheared part. We describe the spacetime of a static and spherically symmetric configuration using the following line element in Schwarzschild coordinates ($t$, $r$, $\theta$, $\phi$):
\begin{align}
ds^2 = -e^{2\nu(r)} dt^2 + e^{2\lambda(r)} dr^2 + r^2 \left( d\theta^2 + \sin^2\theta d\phi^2 \right).    
\end{align}
The function $e^{2\lambda}$ can be rewritten as
\begin{align}
e^{2\lambda} = \left(1 - \frac{2m}{r}\right)^{-1}  \,,  
\end{align}
where $m$ describes the amount of mass contained within a sphere of radius $r$. Meanwhile, the function $\nu$ is related to the gravitational potential. 

The spacetime geometry, described by the above metric, is influenced by the distribution of matter and energy, encapsulated in the stress-energy tensor: 
\begin{align}
    T_{\alpha \beta} = \rho u_\alpha u_\beta + p_r k_{\alpha} k_\beta + p_t \Omega_{\alpha \beta}.
\end{align}
Here $\rho$ is the energy density, $p_r$ is the radial pressure, $p_t$ is the tangential pressure, $u^\alpha$ is the four-velocity, $k^\alpha$ is the unit normal vector in the radial direction orthogonal to $u^\alpha$, and $\Omega_{\alpha \beta} = g_{\alpha\beta}+u_\alpha u_\beta-k_\alpha k_\beta$ is the transverse metric on a 2D sphere orthogonal to both $u^\alpha$ and $k^\alpha$. The unsheared reference state is described by the (flat) material space metric $k_{AB}$ in Sec.~8 of \cite{Karlovini:2002fc}. Hence, the background configuration described by the curved spacetime metric is strained relative to the prescribed reference state due to self-gravity. Note that it is possible for the unsheared reference state to be the same as the static spherically symmetric configuration, depending on the formation process and the material properties\footnote{We remark here that, as stated in \cite{Carter:1972, Finn_1990}, it is likely for the solid to be far from the reference state under the high pressure environment of a NS. On top of that, the reference state may not even exist.} \footnote{This unsheared scenario can be recovered by having a curved material space metric, $e^{\tilde{\lambda}}[1+(1/3)d\ln{n}/d\ln{r}] = e^{{\lambda}} $ (which gives $z \equiv 1$), in Eqs.~(187)-(189) of \cite{Karlovini:2002fc}. 
}

Using the above metric and the stress-energy tensor, one finds, through the Einstein equations, the relativistic hydrostatic equilibrium equations (Tolman-Oppenheimer-Volkoff (TOV) equations) for anisotropic fluids \cite{Bowers_Liang1974}:
\begin{align}
\frac{\mathrm{d} m}{\mathrm{d} r} & =4 \pi r^2 \rho\, ,\label{eq: anisotropic TOV m}\\
\frac{\mathrm{d} p_r}{\mathrm{d} r} & =-\left(\rho+p_r\right) \frac{m+4 \pi r^3 p_r}{r(r-2 m)}-\frac{2 \sigma}{r}\,.
\label{eq: anisotropic TOV pr}
\end{align}
Here, $\sigma$ represents the pressure anisotropy, defined as the difference between radial and tangential pressure:
\begin{align}
    \sigma\equiv p_r-p_t. \label{eq: sigma def}
\end{align}
The stellar configuration reduces to the isotropic one in the $\sigma \to 0$ limit. Within the framework of the quasi-Hookean material defined by Eqs.~\eqref{eq: quasi-Hookean EOS}--\eqref{eq: shear scalar in z}, the anisotropic TOV equations can be reformulated into a system of differential equations involving $m$, $\tilde{p}$, and $z$ as independent variables. The system of differential equations becomes~\cite{Karlovini:2002fc}
\begin{align}
\frac{\mathrm{d} m}{\mathrm{~d} r} &= 4 \pi r^2 \rho\,, \label{eq: m}\\
\frac{\mathrm{d} \tilde{p}}{\mathrm{~d} r} &= \frac{\tilde{\beta}}{r \beta_r}\Bigg\{-\left(\rho + p_r\right) \frac{m + 4 \pi r^3 p_r}{r - 2 m} - 2 \sigma \nonumber\\
    &\quad + 4\left(\mathrm{e}^{\lambda} z - 1\right)\left[\tilde{\mu} + 3 \tilde{\mu} \mathcal{S}^2 - \frac{\sigma}{2}(1 - \tilde{\Omega})\right] \Bigg\} \,, \label{eq: p_check}\\
\frac{\mathrm{d} z}{\mathrm{~d} r} &= \frac{z}{r}\left[\frac{r}{\tilde{\beta}} \frac{\mathrm{d} \tilde{p}}{\mathrm{~d} r} - 3\left(\mathrm{e}^{\lambda} z - 1\right)\right],
\label{eq: z}
\end{align}
where
\begin{align}\centering
& \sigma=-\frac{\tilde{\mu}}{2}\left(z^{-2}-z^2\right), \label{eq: sigam-z}\\
& p_r=\tilde{p}+(\tilde{\Omega}-1)\tilde{\mu} \mathcal{S}^2+\frac{2}{3} \sigma\,\label{eq: pr}, \\
&\tilde{\beta}=(\tilde{\rho}+\tilde{p}) \frac{\mathrm{d} \tilde{p}}{\mathrm{~d} \tilde{\rho}}\,, \quad\tilde{\Omega}=\frac{\tilde{\beta}}{\tilde{\mu}} \frac{\mathrm{d} \tilde{\mu}}{\mathrm{d} \tilde{p}}\,,\\
& \beta_r=\tilde{\beta}+\frac{4}{3} \tilde{\mu}+\left[\tilde{\Omega}(\tilde{\Omega}-1)+\tilde{\beta} \frac{\mathrm{d} \tilde{\Omega}}{\mathrm{d} \tilde{p}}\right] \tilde{\mu} \mathcal{S}^2\nonumber\\
&\qquad\quad+4\left[\tilde{\mu} \mathcal{S}^2+\frac{\sigma}{3}\left(\tilde{\Omega}-\frac{1}{2}\right)\right]. \label{eq: beta_r}
\end{align}
We define $\beta_r$ as the longitudinal modulus in the radial direction, which characterizes the longitudinal wave speeds in that direction.
The above set of equations reduces to the isotropic fluid case when $\tilde{\mu}$ is zero\footnote{The typos mentioned in~\cite{Karlovini_Typo_2005} have been corrected here.}. 
Besides, $z$ always remains less than or equal to 1 because, at the unsheared point (where $\tilde{\mu} = \sigma = 0$ and $ z = 1 $), the derivative $\mathrm{d}z / \mathrm{d}r$ is always negative, which prevents $z$ from exceeding 1. This ensures that the radial pressure can never exceed the tangential pressure, and thus, $\sigma$ always remains negative.

In short, Eqs.~\eqref{eq: m}--\eqref{eq: z} describe the same structural framework as Eqs.~\eqref{eq: anisotropic TOV m}--\eqref{eq: anisotropic TOV pr}, where the thermodynamic quantities are governed by a two-parameter EOS involving the number density and the shear scalar (see Eq.~\eqref{eq: quasi-Hookean EOS}). To fully describe the system, it is necessary to solve for the shear profile, which is determined by the differential equation for $z$. Thus, the results of the structural equations can be recovered by solving the anisotropic TOV equations (Eqs.~\eqref{eq: anisotropic TOV m}--\eqref{eq: anisotropic TOV pr}) if one uses the anisotropy profile computed from the solution for the $z$ equation. Alternatively, one can construct a phenomenological model for $\sigma$, similar to that in~\cite{Bowers_Liang1974} or~\cite{Horvat_2011}, that effectively captures the $\sigma$ profile obtained by solving $z$, and uses that to solve for the anisotropic TOV equations.
In Sec.~\ref{Sec: Anisotropic Profile Modelling}, we further demonstrate how to construct such a phenomenological model for $\sigma$.

\subsection{\label{Sec: BC}Boundary Conditions}

We now discuss the necessary boundary conditions to solve the above set of equations.
Equations~\eqref{eq: m}--\eqref{eq: z} should be integrated from a point relatively close to $r=0$. The initial conditions at this point are as follows \cite{Karlovini:2002fc}:
\begin{align}
& m=\frac{4}{3} \pi \rho_c r^3 + \mathcal{O}(r^4), \label{eq: m initial}\\
& \tilde{p}=p_c-2\pi \tilde{\beta}_c \frac{\left(\rho_c+p_c\right)\left(\rho_c+3 p_c\right)-4 \tilde{\mu}_c \rho_c}{3\left(\tilde{\beta}_c+\frac{4}{3} \tilde{\mu}_c\right)} r^2 + \mathcal{O}(r^3), \\
& z=1-4\pi \frac{\left(\rho_c+p_c\right)\left(\rho_c+3 p_c\right)+3 \tilde{\beta}_c \rho_c}{15\left(\tilde{\beta}_c+\frac{4}{3} \tilde{\mu}_c\right)} r^2 + \mathcal{O}(r^3).
\end{align}
All variables with the subscript $c$ represent values at the center.

The stellar surface is determined by the condition $p_r/p_c < 10^{-14}$. At the surface, the interior solutions are smoothly matched to the Schwarzschild solution, which is valid in the exterior region.

Furthermore, since we are considering an HS with a phase transition, the following junction condition on $\tilde{p}$ is needed to integrate Eq.~\eqref{eq: p_check} from the solid core to the fluid envelope \cite{Karlovini:2002fc}:
\begin{align}
    \tilde{p}_{+}=\tilde{p}_{-}-\frac{\tilde{\mu}_{-}}{6}[(1-&\tilde{\Omega}_{-})(z_{-}^{-1}-z_{-})^2+2(z_{-}^{-2}-z_{-}^2)],\label{eq: Junction Condition}
\end{align}
which is equivalent to the continuity of $p_r$ at the phase transition. The subscripts $+$ and $-$ represent the quantities evaluated at the outer side and the inner side of the interface, respectively. We also have $\tilde{p}_{+} = p_\mathrm{trans}$ for the perfect fluid envelope. This approach ensures that, for cases where $ p_c $ exceeds $ p_\mathrm{trans} $, the star remains an HS (further discussion is provided in Appendix~\ref{AP: junction condition}).

\section{\label{Sec: Stellar Structure}Stellar Structure}
In previous studies, shear deformation has typically been introduced at the perturbative level. However, in this work, we consider a more realistic scenario for HSs with a crystallized core within a fully relativistic elasticity framework, where shear stress is non-vanishing, even at the background (static and spherically symmetric) level.
With the help of structural equations assuming the quasi-Hookean EOS described in Sec.~\ref{Sec: EOS}, we can now examine how the inclusion of anisotropy impacts the results for elastic HSs by comparing them with the typical isotropic models. 

For a better comparison, we use the four elastic HS models (HS-A, HS-B, HS-C, and HS-D) in Fig.~\ref{fig: EOS}, with the same color scheme in other related figures for consistency. We show that the elasticity induces a pressure anisotropy with $\sigma < 0$, which causes an increase in the maximum mass compared to the unsheared models.

\subsection{\label{Sec: Anisotropic profile of the core}Ansitropy Profile of the Core}
\begin{figure}[!htp]
    \centering
    \includegraphics[width=8.6cm]{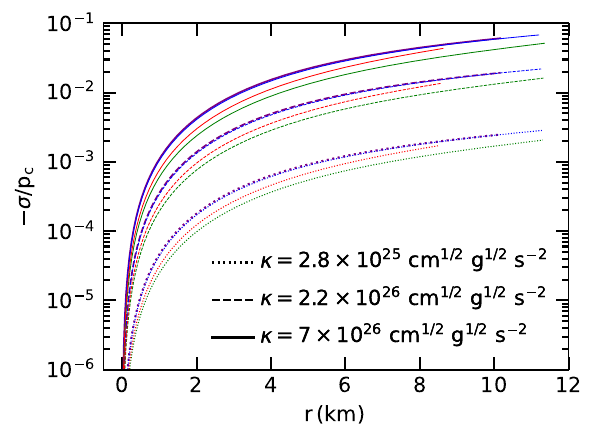}
    \caption{\label{fig: sigma profile} 
The anisotropy profile for elastic HS models (HS-A, HS-B, HS-C, and HS-D) is shown with varying shear modulus coefficients $\kappa$ and a fixed central pressure $p_c = 4.23 \times 10^{35} \, \text{dyn cm}^{-2}$, corresponding to the crosses in Fig.~\ref{fig: MR}. Three different values of $\kappa$ are represented: $(0.28,2.2,7.0)\times10^{26}\text{~cm}^{1/2}\text{~g}^{1/2}\text{~s}^{-2}$ depicted by dotted, dashed, and solid lines, respectively.
    }
\end{figure}

\begin{figure}[!htp]
    \centering
    \includegraphics[width=8.6cm]{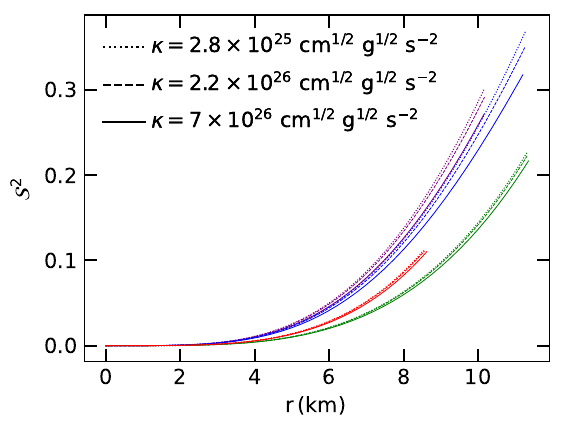}
    \caption{\label{fig: S2-r profile}     
The radial profile of the shear scalar $\mathcal{S}^2$ for elastic HS models, corresponding to the configurations in Fig.~\ref{fig: sigma profile}.
   }
\end{figure}

To visualize how anisotropy manifests within the core of the star, we present the anisotropy profile for elastic HS models with different shear modulus coefficients $\kappa$ in Fig.~\ref{fig: sigma profile}. At the stellar center, both radial and tangential pressures are effectively equal due to the symmetric environment and minimal deformation, resulting in vanishing anisotropy. Moving outward from the core, the symmetry is gradually broken as shear strain develops, as indicated by $\mathcal{S}^2$ in Fig.~\ref{fig: S2-r profile}. This leads to the emergence of pressure anisotropy, which monotonically increases in magnitude with radius.

Comparing the same model with different shear modulus coefficients $\kappa$ (as seen in the same colored lines with different line styles in Fig.~\ref{fig: sigma profile}), we observe that for larger values of $\kappa$, the anisotropy increases more significantly with increasing radius. Since $\sigma \propto \kappa$ (from Eq.~\eqref{eq: sigam-z}), a more rigid material (with a higher $\kappa$) leads to a more pronounced increase in anisotropy because the material can resist shear deformation more effectively.

\begin{figure}[!htp]
    \centering
    \includegraphics[width=8.6cm]{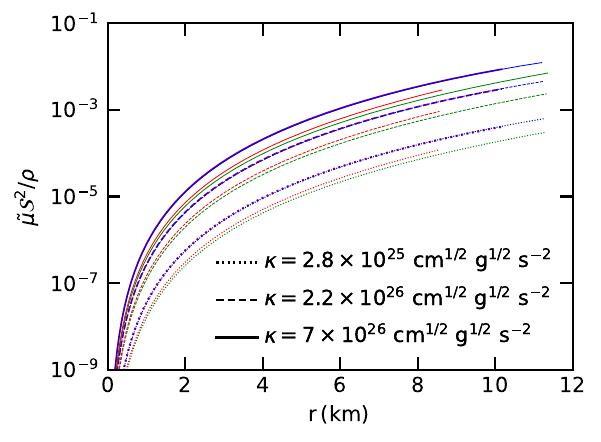}
    \caption{\label{fig: shear domain} 
    The radial profile of the contribution of the shear term $\tilde{\mu}\mathcal{S}^2$ on the energy density of elastic HS models corresponding to the configurations in Fig.~\ref{fig: sigma profile}.
    }
\end{figure}

In comparing different EOS models (as shown by the different colored lines with the same line style in Fig.~\ref{fig: sigma profile}), we observe that the energy density gap has a minimal impact on the anisotropy profile because it does not affect the magnitude of the shear part contribution of these models, as the blue (HS-A) and purple (HS-C) models demonstrated in Fig.~\ref{fig: shear domain}. 
In contrast, the fluid sound speed $\tilde{v}$ associated with the unsheared part plays a more significant role in determining the anisotropic behavior (the purple curves have a smaller $\tilde{v}$ compared to the green curves).
Since the overall anisotropy arises from contributions of the sheared term, if the unsheared part dominates, the resulting anisotropy will be smaller as the shear strain is reduced in such configurations.

The sheared contribution to the energy density is shown in Fig.~\ref{fig: shear domain}. The results illustrate a trend of increasing the shear contribution with the radius, which aligns with the anisotropy profile discussed before. As the radius increases, the shear scalar $\mathcal{S}^2$ increases, leading to a progressively larger contribution of the shear term to the overall energy density. This trend highlights the growing role of shear deformation in the outer regions of the elastic core. Furthermore, the magnitude of the shear contribution is strongly dependent on the shear modulus coefficient $\kappa$. For larger values of $\kappa$, the shear term contribution to the energy density is significantly enhanced. This behavior reflects the higher rigidity of the material, allowing it to maintain greater shear stress and deformation, thus amplifying its impact on the energy density profile.

\subsection{\label{Effects on the macroscopic properties}Effects on the Macroscopic Properties}
We next examine how pressure anisotropy $\sigma$ influences HS properties, such as mass, radius, and compactness, providing a comprehensive understanding of how anisotropy alters the behavior of elastic HSs compared to isotropic models.

In Fig.~\ref{fig: MR}, we show the effects of different shear moduli on the mass and radius of elastic HSs. We observe that the maximum mass of elastic HS increases with $\kappa$. Among the three groups shown, the maximum mass of the models with the largest $\kappa$ increases by up to 5.242\% (HS-A, blue), 3.263\% (HS-B, green), 4.648\% (HS-C, purple), 1.844\% (HS-D, red).
This conclusion differs from the findings presented in \cite{Karlovini:2002fc}. A detailed discussion of this discrepancy is provided in Appendix~\ref{AP: karlovini}.
We also observe that the differences in the $M$--$R$ curves arise from variations in $\tilde{v}$ and $\Delta \tilde{\rho}$ (A similar analysis can also be found in \cite{Dong2024}, although it is limited to isotropic HSs.). The purple curves have a smaller $\tilde{v}$ compared to the green curves, meaning that the material in the purple model is less resistant to pressure change. This reduced stiffness results in smaller radii and lower maximum masses, shifting the $M$--$R$ curve down and to the left. Meanwhile, the blue models have a smaller energy density gap than the purple models, which increases the stability of higher-mass configurations by making it favorable for the star to support greater masses. This effect shifts the blue $M$--$R$ curve up and to the right, allowing for higher maximum masses and larger radii compared to the purple curves.

Some observational constraints are overlaid on the $M$--$R$ relations, including the mass measurement of PSR J0740+6620~\cite{Cromartie_2020}, as well as the inferred mass and radius constraints from the binary NS merger event GW170817~\cite{Abbott_2018} (see similar constraint in~\cite{De2018, Zhao2018}), and the latest constraints from NICER observations of PSR J0030+0451~\cite{Vinciguerra_2024}, PSR J0740+6620 (combined with XMM-Newton data)~\cite{Salmi_2024}, and PSR J0437--4715~\cite{Choudhury_2024}. 
Due to the increase in the maximum mass caused by the elasticity, we can expect that some stars that previously could not satisfy the mass constraints of J0740+6620 may become viable. However, the effects of elasticity cannot be distinguished by current observational constraints. For example, the HS-A model can satisfy all constraints, regardless of whether elasticity is taken into account or not.

\begin{figure}[!htp]
    \centering
    \includegraphics[width=8.6cm]{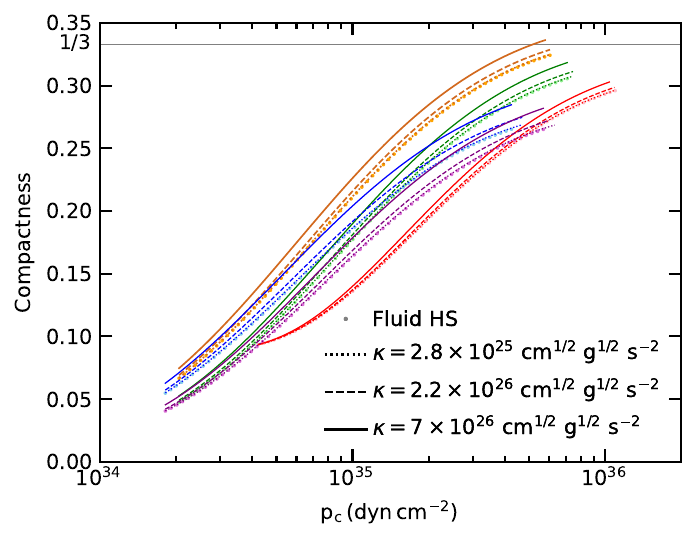}
    \caption{\label{fig: PC} The relation between compactness and central pressure with different shear modulus (represented by $\kappa$) for elastic HSs (HS-A, HS-B, HS-C, and HS-D; same color coding as in Fig.~\ref{fig: EOS}). An additional model is shown in brown, with EOS parameters: ($p_\mathrm{trans}$, $\tilde{v}^2$, $\Delta \tilde{\rho}$)= ($4.89 \times 10^{33}\,\text{dyn cm}^{-2}$, 0.75, $ 8.15 \times 10^{13}\,\text{g cm}^{-3}$). This model satisfies all the constraints discussed in Sec.~\ref{Sec: EOS}.
    The curves show the full stable branch up to the maximum-mass configurations. For comparison, the light-color dots represent the isotropic fluid HSs. The grey line indicates the stability boundary for the star. }
\end{figure}

As demonstrated in~\cite{Alho_2022}, elastic stars can have high compactness while maintaining radial stability due to anisotropy. A stable configuration with compactness $> 1/3$ can potentially be a black hole mimicker, i.e., a horizonless object with a photon sphere. Here, we observe a similar behavior with the particular quasi-Hookean EOS (Eqs.~\eqref{eq: quasi-Hookean EOS}--\eqref{eq: shear scalar}) rather than the extended polytropic EOS\footnote{A generalization of the fluid polytropic EOS to include elasticity by adding a term quadratic in shear deformation (see \cite{Alho_2022, Alho_2024}).} in \cite{Alho_2022}.
Figure \ref{fig: PC} illustrates the relation between compactness and central pressure under different shear moduli for various elastic HSs. We observed that the compactness of the elastic HS is greater than that of the fluid HS, and this feature becomes more significant as the central pressure increases. For a fixed central pressure, the compactness increases as one increases $\kappa$. For the maximum mass star, the increase in compactness is 5.79\% (HS-A, blue), 3.46\% (HS-B, green), 4.73\% (HS-C, purple), 2.08\% (HS-D, red), respectively. 

Additionally, we present results for an extra set of HS models (brown) in Fig.~\ref{fig: PC}, demonstrating that with the inclusion of anisotropy, the stellar compactness can approach or even exceed 1/3 before reaching the maximum mass, resulting in ultra-compact HSs~\cite{Cunha_2017}. 
It is expected that these ultra-compact stars within the unstable light ring ($R < 3M$) possess a second stable light ring with a smaller radius. This causes their perturbations to be unstable either in the nonlinear regime or at the linear level with small rotations, suggesting that the ultra-compact star will eventually become a black hole or a compact star without a photon sphere \cite{Cardoso_2014, Keir_2016}. However, these results are obtained by analyzing geodesics. The actual fate of these stars requires studying the elastodynamics in the perturbed configurations, which is beyond the scope of this work.

\section{Anisotropic Profile Modelling}\label{Sec: Anisotropic Profile Modelling}
Next, we construct a new phenomenological model for pressure anisotropy.
Our parameterized model (Eq.~\eqref{eq: sigma fit}) is a function of both the thermodynamic variables and the local curvature. The model is derived by fitting the numerical anisotropy profiles in Sec.~\ref{Sec: Stellar Structure}, obtained with the quasi-Hookean EOS that guarantees the $\sigma$ to vanish at $r = 0$. 
We stress that Eq.~\eqref{eq: sigma fit} is valid in the core region, and the core surface is determined by the phase transition between the core and the envelope, as detailed in Sec.~\ref{Sec: BC}. 

Notice the following properties for $\sigma_\mathrm{fit}$:
(i) At the stellar center, the equality between $p_r$ and $p_t$ is automatically satisfied. This is because the mass function $m(r)$ behaves as $m\sim r^3$ near $r=0$ (see Eq.~\eqref{eq: m initial}), ensuring that $(1-2m/r)^{-1}$ remains well-behaved and does not introduce any singularities. Moreover, thanks to the quasi-Hookean EOS, the energy density $\rho$ and pressure $p$ are expressed as the sum of a relaxed state and a term accounting for shear deformation. This formulation leads to the absence of shear at the stellar center. (ii) $\sigma_\mathrm{fit}$ is discontinuous at the interface as it remains finite there while $\sigma=0$ in the outer fluid envelope.

Let us further analyze the implicit relation between $\lambda$ and $N$ with respect to $p_c$ and $\kappa$. These parameters obviously depend on $\kappa$ because the shear modulus clearly impacts the stellar structure, as discussed in the previous section. They also depend on $p_c$ because the solution to the set of structural equations depends on this quantity through the boundary condition at the center.

\begin{figure}[!htb] 
    \centering
    \includegraphics[width=8.6cm]{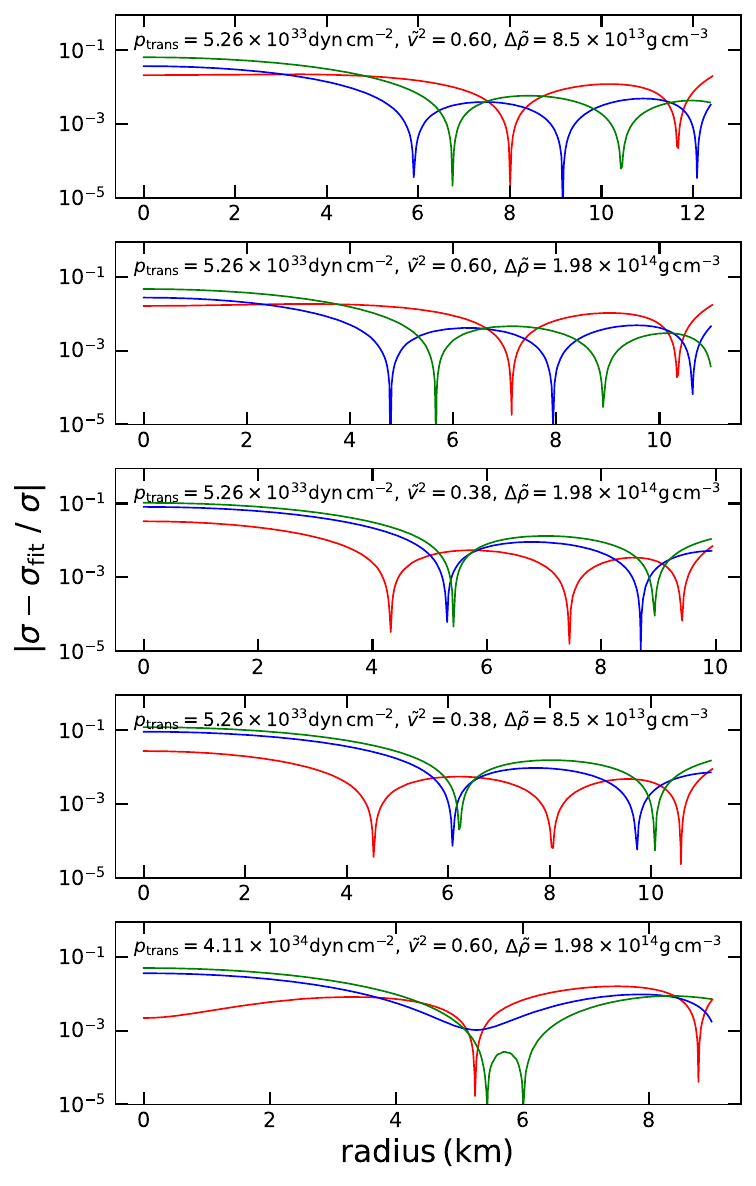}
    \caption{Fractional difference between numerical results and the fitting formula (Eq.~\eqref{eq: sigma fit}) for HSs at the maximum mass with three different values of $\kappa$: $2.8 \times 10^{25}\text{cm}^{1/2}\text{g}^{1/2}\text{s}^{-2}$ (green), $2.2 \times 10^{26}\text{cm}^{1/2}\text{g}^{1/2}\text{s}^{-2}$ (blue), and $7 \times 10^{26}\text{cm}^{1/2}\text{g}^{1/2}\text{s}^{-2}$ (red). Each panel corresponds to a different combination of $p_\text{trans}$, $\tilde{v}^2$, and $\Delta\tilde{\rho}$, as labeled in each plot. The first panel adopts a new EOS model that satisfies all the constraints discussed in Sec.~\ref{Sec: EOS}. The second to fifth panels correspond to HS-B, HS-C, HS-A, and HS-D, respectively. } 
    \label{fig: fitting} 
\end{figure}

For each specific HS model with defined parameters ($p_\mathrm{trans}$, $\tilde{v}$, $\Delta \tilde{\rho}$),
we also need to determine the ranges of $p_c$ and $\kappa$ to complete the fitting. For example, for the results plotted in this work, we evaluate a range of $p_c$ from $1.82 \times 10^{34}$ dyn cm$^{-2}$ up to the value corresponding to the maximum mass, which depends on the specific model. The shear modulus parameter $\kappa$ is varied from $2.8 \times 10^{25} \, \mathrm{cm}^{1/2} \, \mathrm{g}^{1/2} \, \mathrm{s}^{-2}$ to $7 \times 10^{26} \, \mathrm{cm}^{1/2} \, \mathrm{g}^{1/2} \, \mathrm{s}^{-2}$.
We construct the relation in two steps, as explained below:
\begin{enumerate}
    \item For a specific HS model with a fixed $\kappa$, fit the functional forms of $\lambda$ and $N$ using various $\bar{p}_c=p_c/(1 \, \mathrm{dyn}~\mathrm{cm}^{-2})$. The result is given as follows, where $\mathcal{Y}$ symbolizes $-\lambda$ or $-N$:
    \begin{align}
    \ln(\mathcal{Y}) = \sum_{i=0}^{4} a_i [\ln(\bar{p}_c)]^i. \label{eq: PC fit}
    \end{align}
    Here, each coefficient $a_i$ is a fitting parameter but is also a function of $\kappa$.
    \vspace{1pt}
    \item Fit the coefficients $a_i$ ($i=0,1,2,3,4$) with the corresponding $\kappa$. Note that $\bar{\kappa}$ is defined as a dimensionless quantity, $\bar{\kappa} = \kappa / (10^{26} \, \mathrm{cm}^{1/2} \, \mathrm{g}^{1/2} \, \mathrm{s}^{-2})$. The coefficients can be expressed as follows:
    \begin{align}
    a_i &= \sum_{j=0}^{4} b_j \left[\bar{\kappa}\right]^j + \sum_{l=1}^{4} c_l \left[\bar{\kappa}\right]^{-l} . \label{eq: kappa fit}
    \end{align}
    Again, all $b_j$ and $c_l$ are fitting parameters for a specific $a_i$ in Eq.~\eqref{eq: PC fit}.
\end{enumerate}
In Appendix~\ref{AP: fitting example}, we provide tables of fitted parameters for the HS-A model as an example. We also provide a Mathematica notebook for this fitting process with examples shown in Fig.~\ref{fig: fitting} in the GitHub repository~\cite{anisotropy_fitting_process}.

Figure~\ref{fig: fitting} presents the fractional differences between numerical results and our parameterized model of the anisotropy for different HS models. We observe that the error is maintained below 10\%.\footnote{In addition to the models shown in Fig.~\ref{fig: fitting}, we have also tested cases with a lower transition pressure of $2 \times 10^{33}\,\text{dyn cm}^{-2}$, which satisfy the causality and observational constraints. The fitting error in these cases remains below 10\% as well.} This indicates that our new analytical form of $\sigma$ is valid across a broad parameter space.

While the new model achieves agreement with the numerical results, it is still necessary to compare it with other commonly used anisotropy models, such as the BL model~\cite{Bowers_Liang1974} and the H model~\cite{Horvat_2011} (see Appendix~\ref{AP: fitting example} for more details on these models). Figure~\ref{fig: sigma compares} illustrates that these widely used models are not suitable for describing the anisotropic behavior of the elastic core in HSs because of their significant deviations from the realistic profile.

\section{\label{Sec: Conclusion}Conclusion}
In this work, we studied the effects of nonlinear elasticity on the static, spherically symmetric structures of HSs with a solid core. The most important distinction of this work compared to our previous work~\cite{Dong2024} lies in the inclusion of shear deformation into the background structure. The formulation is based on the fully relativistic nonlinear elasticity theory in \cite{Carter:1972}. We modeled the QM core as a solid described by the quasi-Hookean EOS (Eq.~\eqref{eq: quasi-Hookean EOS}), where the unsheared component follows the CSS template (Eq.~\eqref{eq: EOS}), and the sheared component is modeled by the shear scalar defined in Eq.~\eqref{eq: shear scalar}, originally proposed in \cite{Karlovini:2002fc}. The NM envelope, meanwhile, is treated as a perfect fluid layer described by the APR EOS.

Next, we examined the anisotropic profiles of HSs to understand the qualitative effects of elasticity. Shear stress introduces anisotropy in the pressure distribution within the core. Specifically, the tangential pressure $p_t$ consistently exceeds the radial pressure $p_r$, indicating an asymmetry driven by the shear deformation. At the center of the star, anisotropy vanishes because of the symmetric environment and the minimal deformation. However, the magnitude of anisotropy, $|\sigma|$, increases monotonically with the radius from the core’s center. This trend highlights the critical role of shear deformation in determining the anisotropy profiles of elastic HSs.

We found that the anisotropy significantly affects the maximum mass and compactness of HSs. For stars with a large shear modulus, the maximum mass increases by around 10\%, and the compactness shows a marked enhancement compared to isotropic models. Under the current observational constraint, the increased maximum mass allows for the inclusion of softer EOSs. However, it remains unrealistic to distinguish whether elastic HSs exist in the stellar interior with the current observations, as the effects of elasticity are mainly concentrated in the high-mass region. 
Additionally, for some EOS parameters, the HS models along the $M$--$R$ relation before reaching the maximum mass can have compactness exceeding $1/3$. These objects can be black hole mimickers that are stable against linear radial perturbations if the turning point of the $M$--$R$ curve implies a change in radial stability as in the homogeneous fluid models \cite{Bardeen_1966, Shapiro_1983}. This has been shown to hold for a solid star following the quasi-Hookean EOS, but may not apply to a star with a density discontinuity\footnote{It is known that for HS models with a sharp phase transition, the turning point of the $M$--$R$ curve does not necessarily imply a change in stability even for fluid stars. Depending on the rate of the phase conversion, there can be an extended stable region passing the turning point (see, e.g., \cite{Karlovini_2004, Pereira_2018}).} \cite{Karlovini_2004}. Even being stable against linear radial pulsations, whether these ultra-compact HSs can exist without being prone to nonlinear instabilities still requires detailed perturbation studies involving elastodynamics. Therefore, shear effects play a critical role in refining EOS models and accurately capturing the structural properties of HSs.

Lastly, we show that phenomenological anisotropy models that are commonly used in literature (BL and H models) cannot accurately capture the anisotropy profile for elastic HS cores and propose a new parametrized model of anisotropy that can capture the realistic profile with an error of $\sim$10\%. This new model bridges the gap between microphysical EOSs and parametrized anisotropic models, offering a practical tool for exploring the effects of elasticity in astrophysical contexts. 

We note that the estimate of shear modulus (Eq.~\eqref{eq: shear modulus}) is based on an idealized ultrarelativistic approximation, and deviations may arise in more realistic scenarios, particularly at lower densities where corrections to the scaling relations become significant. While this may lead to some overestimation of the shear modulus, the qualitative impact of elasticity on the increase in the maximum mass remains robust, as the additional pressure support from anisotropy persists even with moderate variations in $\kappa$.

Future work can focus on extending this model to study nonradial perturbations using the quasi-Hookean framework, including the formulation and calculation of tidal deformability and pulsation modes. 
Our subsequent work~\cite{Yip2025} will explore the potential universality of HS properties by using this quasi-Hookean EOS. One can also compare the quasi-Hookean approach with other elastic EOSs, such as those recently discussed in \cite{Alho_2024}.

\begin{acknowledgments}
The authors thank the anonymous reviewers for their valuable suggestions to help improve this manuscript.
They also thank Chun-Ming Yip for helping us check some of our calculations. Z. Z. Dong thanks Sophia Han for the valuable discussions, as well as to the attendees of \textit{Compact Stars in the QCD Phase Diagram 2024} at YITP, Kyoto University, for their helpful comments.
K.Y. acknowledges support from NSF Grant PHYS-2339969 and the Owens Family Foundation.

\end{acknowledgments}

\appendix
\section{\label{AP: fitting example} Examples for $\sigma$ model}
In this appendix, we provide some examples of anisotropic models. First, we list two phenomenological models that are commonly used in literature that we also use in Fig.~\ref{fig: sigma compares}. The first model is the H model, originally introduced by Horvat \textit{et al.} \cite{Horvat_2011}, given by
\begin{align}
    \sigma=2 \lambda_{\mathrm{H}} p_r \frac{m}{r}.
\end{align}
The second model is the BL model, originally proposed by Bowers and Liang \cite{Bowers_Liang1974} (the BL model), defined by
\begin{align}
    \sigma = \frac{\lambda_{\mathrm{BL}}}{3}r^2(\rho+3 p_r)(\rho+p_r)\left(1-\frac{2 m}{r}\right)^{-1}.
\end{align}
For both models, the constant parameters $\lambda_{\mathrm{H}}$ and $\lambda_{\mathrm{BL}}$
quantify the degree of anisotropy with $\lambda_{\mathrm{H}} \to 0$ or $\lambda_{\mathrm{BL}} \to 0$ corresponding to the isotropic limit. 

Next, we present a set of fitted parameters for our new phenomenological anisotropy model using the HS-A model in Tables~\ref{tab:parameters_part1} \&~\ref{tab:parameters_part2}. This fitting is applicable across the full range of $\kappa$ and $p_c$ for the stable star. A Mathematica notebook detailing the fitting process for this model has been made available at the GitHub repository~\cite{anisotropy_fitting_process}, allowing readers to better understand our method and generate parameters using some input values of $p_\mathrm{trans}$, $\tilde{v}^2$, and $\Delta \tilde{\rho}$.

\begin{table*}[htbp]
\centering
\begin{tabular}{cccccccccc}
\toprule
 & $b_0$ & $b_1$ & $b_2$ & $b_3$ & $b_4$ & $c_1$ & $c_2$ & $c_3$ & $c_4$ \\
\midrule
$a_0$ &$-3.1536E+1$ & $9.2881E+0$ & $-1.7453E+0$ &  $1.3030E-1$ & $-2.4949E-3$ & $3.1203E+1$ & $-1.8807E+1$ & $5.1607E+0$ & $-5.1826E-1$ \\
$a_1$ &$-4.7791E+0$ & $1.4354E+0$ & $-2.7052E-1$ &  $2.0186E-2$ & $-3.8309E-4$ & $4.8586E+0$ & $-2.9270E+0$ & $8.0258E-1$ & $-8.0535E-2$ \\
$a_2$ &$-3.2598E-1$ & $9.8503E-2$ & $-1.8581E-2$ &  $1.3856E-3$ & $-2.6089E-5$ & $3.3500E-1$ & $-2.0172E-1$ & $5.5276E-2$ & $-5.5429E-3$ \\
$a_3$ &$-8.2378E-3$ & $2.5308E-3$ & $-4.7728E-4$ &  $3.5577E-5$ & $-6.6567E-7$ & $8.6312E-3$ & $-5.1953E-3$ & $1.4228E-3$ & $-1.4258E-4$ \\
$a_4$ &$-8.2733E+1$ & $2.3003E+1$ & $-4.2722E+0$ &  $3.1978E-1$ & $-6.2801E-3$ & $7.3731E+1$ & $-4.4808E+1$ & $1.2333E+1$ & $-1.2407E+0$ \\
\bottomrule
\end{tabular}
\caption{Fitted coefficients in Eq.~\eqref{eq: kappa fit} for 
$\mathcal{Y} = -\lambda$ in Eq.~\eqref{eq: PC fit} for the HS-A model.}
\label{tab:parameters_part1}
\end{table*}

\begin{table*}[htbp]
\centering
\begin{tabular}{cccccccccc}
\toprule
& $b_0$ & $b_1$ & $b_2$ & $b_3$ & $b_4$ & $c_1$ & $c_2$ & $c_3$ & $c_4$ \\
\midrule
$a_0$ &$-3.3506E+1$ & $1.1016E+1$ & $-1.5123E+0$ &  $1.4207E-2$ &  $6.9305E-3$ & $4.6180E+1$ & $-2.8529E+1$ & $7.8738E+0$ & $-7.8857E-1$ \\
$a_1$ &$-5.0451E+0$ & $1.6790E+0$ & $-2.2787E-1$ &  $1.2972E-3$ &  $1.1251E-3$ & $7.1472E+0$ & $-4.4151E+0$ & $1.2176E+0$ & $-1.2180E-1$ \\
$a_2$ &$-3.4489E-1$ & $1.1374E-1$ & $-1.5252E-2$ &  $3.3850E-5$ &  $8.0348E-5$ & $4.8998E-1$ & $-3.0267E-1$ & $8.3406E-2$ & $-8.3350E-3$ \\
$a_3$ &$-8.7874E-3$ & $2.8886E-3$ & $-3.8285E-4$ & $-3.3658E-7$ &  $2.1296E-6$ & $1.2558E-2$ & $-7.7565E-3$ & $2.1361E-3$ & $-2.1326E-4$ \\
$a_4$ &$-8.8579E+1$ & $2.7606E+1$ & $-3.8210E+0$ &  $5.4904E-2$ &  $1.5647E-2$ & $1.1036E+2$ & $-6.8545E+1$ & $1.8963E+1$ & $-1.9027E+0$ \\
\bottomrule
\end{tabular}
\caption{Similar to Table~\ref{tab:parameters_part1} but for $\mathcal{Y} = -N$.}
\label{tab:parameters_part2}
\end{table*}

\section{\label{Sec: Wave speed}Wave Speeds}
The fluid sound speed is constrained by the causality limit, meaning that it cannot exceed the speed of light ($\tilde{v} \leq 1$). However, in elastic materials, the situation becomes more complex due to the presence of multiple wave speeds associated with different modes of deformation. To ensure the physical validity of the model, we must ensure that all of these wave speeds remain below the speed of light. In this appendix, we provide an approximate constraint on the fluid sound speed using Eq.~\eqref{eq: fluid sound speed limit} below and present a numerical analysis of wave speeds, resulting in a parameter space depiction of regions where causality is satisfied (or broken) across various HS models.

\subsection{\label{Sec: Wave speed in elastic matter}Wave Speeds in Elastic Matter}
Within isotropic elastic matter in the linear (Hookean) regime \cite{Landau1986theory}, there are two distinct wave speeds corresponding to the longitudinal wave and transverse wave, respectively. The wave speeds are given by 
\begin{align}
    v_\parallel^2 &= \tilde v^2 + \frac{4}{3}\frac{\tilde \mu}{\tilde \rho + \tilde p}, \\
    v_\perp^2 &= \frac{\tilde \mu}{\tilde \rho + \tilde p},
\end{align}
where $v_\parallel$ is the longitudinal wave speed and $v_\perp$ is the transverse wave speed. Due to the fact that $\tilde{\mu}$ is proportional to $\sqrt{\tilde{\rho}}$, it is straightforward to deduce that both $v_\parallel$ and $v_\perp$ reach their maximum values at the core surface.
We use this expression to establish an approximate upper bound for the fluid sound speed within the solid cores of HSs:
\begin{align}   
\tilde{v}^2 \leq 1 - \frac{4}{3} \frac{\kappa\sqrt{\tilde{\rho}_{\mathrm{NM}}\left(p_\mathrm{trans}\right)+\Delta \tilde{\rho}}} {\tilde{\rho}_{\mathrm{NM}}\left(p_\mathrm{trans}\right)+\Delta \tilde{\rho}+ p_\mathrm{trans}}\,. \label{eq: fluid sound speed limit}
\end{align}
For the specific EOS model, we expect that the upper bound of fluid sound speed will reach a minimum limit when $\kappa$ is at its maximum value. This behavior will be discussed later. 

Moreover, due to the anisotropy, the above two wave modes for isotropic elastic matter exhibit directional dependence as we move away from the center, resulting in the emergence of five distinct wave speeds (see also Eq.~(159) in ~\cite{Karlovini:2002fc}): 
\begin{align}
    v_{r\parallel}^2 &= \frac{\beta_r}{\rho + p_r},\notag \\
    v_{t\parallel}^2 &= \frac{\beta_t}{\rho + p_t}, \notag\\
    v_{r\perp}^2 &= \frac{\tilde{\mu} + \frac{3}{2} (\tilde{\mu}\mathcal{S}^2 -\frac{\sigma}{3})}{\rho + p_t},\label{eq: wave speed} \\
    v_{t\perp r}^2 &= \frac{\tilde{\mu} + \frac{3}{2} (\tilde{\mu}\mathcal{S}^2 +\frac{\sigma}{3})}{\rho + p_r}, \notag\\
    v_{t\perp t}^2 &= \frac{\tilde{\mu} (1+\mathcal{S}^2)}{\rho + p_t},\notag
\end{align}
where $\beta_t$ is the tangential longitudinal modulus
\begin{align}
    &\beta_t=\tilde{\beta}+\frac{4}{3} \tilde{\mu}+\left[\tilde{\Omega}(\tilde{\Omega}-1)+\tilde{\beta} \frac{\mathrm{d} \tilde{\Omega}}{\mathrm{d} \tilde{p}}\right] \tilde{\mu} \mathcal{S}^2 \nonumber\\
    &\qquad\quad+ 2 \left[ \tilde{\mu}\mathcal{S}^2 -\frac{\sigma}{3} \left( \tilde{\Omega} - \frac{1}{2} \right) \right].
\end{align}
Here, both $\beta_r$ (see Eq.~\eqref{eq: beta_r}) and $\beta_t$ reduce to $\tilde{\beta} + 4\tilde{\mu}/3$ in linear elasticity regime (i.e., $z \rightarrow 1$). The longitudinal speeds $v_{r\parallel}$ and $v_{t\parallel}$ describe the speeds of disturbance oscillating parallel to propagation axes along the radial and tangential directions, respectively. Meanwhile, $v_{r\perp}$, $v_{t\perp r}$, and $v_{t\perp t}$ represent the speeds of transverse waves. Specifically, $v_{r\perp}$ is the transverse wave speed propagating along the radial direction, with disturbances perpendicular to the propagation direction (radial). Similarly, $v_{t\perp r}$ is the transverse wave speed in the tangential direction, with disturbances perpendicular to the propagation direction (tangential), influenced by both $p_r$ and $p_t$. In contrast, $v_{t\perp t}$ is the transverse wave speed within the tangential plane, where both the wave propagation and the disturbance lie, and it is determined only by $p_t$. Therefore, we must carefully select the fluid sound speed so that all the wave speeds do not exceed the speed of light, as required by the causality limit in relativistic contexts.

\subsection{\label{Sec: Wave Speed parameter space depiction} Parameter Space Depiction of Wave Speed}
Let us first examine the behavior of wave speeds. In this section, we select a model with a relatively low transition pressure $p_\mathrm{trans} = 2 \times 10^{33}\,\text{dyn cm}^{-2}$, which results in a larger elastic core, enhancing the visibility of anisotropy effects.

\begin{figure}[!htp]
    \centering
    \includegraphics[width=8.6cm]{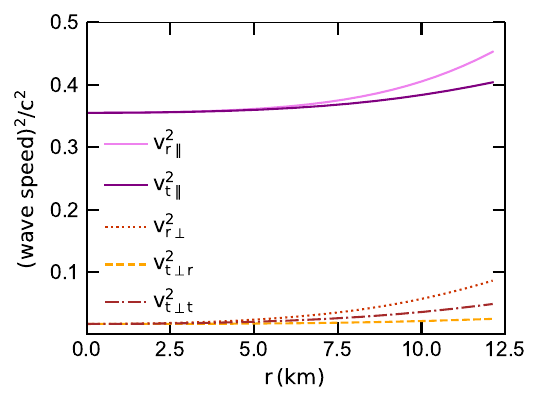}
    \caption{\label{fig: wave speed} 
   The principal speeds in Eq.~\eqref{eq: wave speed} for the maximum mass star of the model with $\kappa=7 \times 10^{26}\text{cm}^{1/2}\text{g}^{1/2}\text{s}^{-2}$. The EOS parameters for this model are ($p_\mathrm{trans}$, $\tilde{v}^2$, $\Delta \tilde{\rho}$) = ($2 \times 10^{33}\,\text{dyn cm}^{-2}$, 0.33, $ 10^{14}\,\text{g cm}^{-3}$). The corresponding mass–radius relation satisfies all current observational constraints mentioned in Fig.~\ref{fig: MR}, ensuring the physical relevance of this model.}
\end{figure}

Figure~\ref{fig: wave speed} illustrates how the wave speeds vary as one moves outward from the core of the star. At the stellar center, where the material is unsheared, we observe only two distinct wave speeds: longitudinal and transverse waves. As the radius increases, all wave speeds show a general upward trend, reaching their maximum values at the core surface, consistent with our previous discussion. This growth in wave speed results from the combined effects of the unsheared and sheared properties of the material. 

\begin{figure}[!htp]
    \centering
    \includegraphics[width=8.6cm]{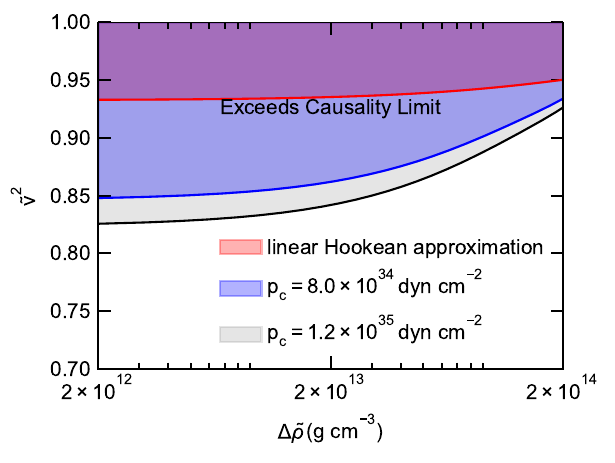}
    \caption{\label{fig: wave speed map} 
    The upper bound of $\tilde{v}^2$ for HS models against energy density gaps, with fixed parameters $\kappa = 7 \times 10^{26}\text{~cm}^{1/2}\text{~g}^{1/2}\text{~s}^{-2}$, $p_\mathrm{trans} = 2 \times 10^{33}\text{~dyn}\text{~cm}^{-2}$. The gray and blue shaded regions correspond to different central pressures, $p_c$, with $1.2 \times 10^{35} \text{ dyn cm}^{-2}$ and $8.0 \times 10^{34} \text{ dyn cm}^{-2}$, respectively. The red-shaded region represents the linear Hookean approximation from Eq.~\eqref{eq: fluid sound speed limit}.
    }
\end{figure}

The wave speed at the core surface is required not to exceed the speed of light. In Fig.~\ref{fig: wave speed map}, we constructed a parameter space depiction of the valid region for fluid sound speed by constraining all the wave speeds in Eq.~\eqref{eq: wave speed}. Models with a large energy density gap can accommodate higher fluid sound speeds, allowing for greater $\tilde{v}^2$ without exceeding the causality constraint. Conversely, for models with a smaller energy density gap, the wave speed is more significantly influenced by the shear component, indicating that the transverse wave component becomes dominant in these scenarios (we discuss this in Sec.~\ref{Sec: Anisotropic profile of the core}). We also observe that the valid region is dependent on the central pressure. For cases with higher $p_c$, the permissible range for wave speeds becomes narrower. We identify a value of 0.8 for $\tilde{v}^2$ as a practical upper limit for all the stable HSs. 

Additionally, we provide in Fig.~\ref{fig: wave speed map} the excluded parameter space computed from a linear Hookean approximation in Eq.~\eqref{eq: fluid sound speed limit}. Notice that the approximation is only valid when the energy density gap is large. In other words, the nonlinearity in the quasi-Hookean relation becomes more important for the low-density gap, which is consistent with the results in Fig.~\ref{fig: shear domain}. As shown in the figure, the green curve represents a smaller density gap compared to the blue curve, leading to a larger core and a larger shear contribution.

\section{\label{AP: junction condition}Junction Condition}

In this appendix, we discuss the junction conditions at the quark-hadron interface. These conditions, in general, depend on the microscopic properties of the phase transition. Throughout this paper, we assume a first-order phase transition, where the solid quark phase has a radial pressure $p_r = p_{\rm trans}$ and the fluid hadronic phase has $p = p_{\rm trans}$ at the phase separation point, following \cite{Karlovini:2002fc}. We show the steps to derive the junction conditions (\cite{Israel_1966}, see also \cite{Poisson_2004} for reference), requiring the continuities of the intrinsic and extrinsic curvature. Note that depending on the phase transition physics, the extrinsic curvature can have a discontinuity, giving rise to new surface degrees of freedom \cite{Pereira_2014, Pereira_2019}. We will discuss this possibility at the end of this appendix.

We use the bracket notation $[\cdot]$ to represent the jump (discontinuity) of a quantity across the hypersurface, defined as

\begin{equation}
[A] = A_{+} - A_{-},
\end{equation}
where $ A_{+} $ and $ A_{-} $ are the values of $ A $ on the exterior and interior sides of the hypersurface, respectively. The hypersurface of the solid-fluid interface is defined as the constant pressure surface with $p_r = p_\mathrm{trans}$, specified by the unit vector $\bm{n} = e^{-\lambda}\partial_r$.

The first condition requires the continuity of the induced metric across the interface,
\begin{equation}
    [h_{\mu \nu}] = 0,
\end{equation}
where $h_{\mu \nu} = g_{\alpha\beta} \mathcal{P}^\alpha_{\;\mu} \mathcal{P}^\beta_{\;\nu}$ is the intrinsic curvature on the hypersurface, and $\mathcal{P}^\alpha_{\;\mu} = \delta^\alpha_{\; \mu} - n^\alpha n_\mu$ is the projection tensor onto the hypersurface. This ensures 
\begin{equation}
    [\nu]=0,\quad [r]=0, \label{eq:junction_1_BC1}
\end{equation}
where the temporal metric component $\nu$ remains continuous across the stellar surface and a smooth matching of the radial coordinate $r$ is assumed for a consistent geometric transition between the interior and exterior spacetime.

The second condition involves the extrinsic curvature $ K_{\mu \nu}$, which encodes how the surface is embedded in the surrounding spacetime, expressed as
\begin{equation}
    [K_{\mu \nu}] = 0,
\end{equation}
where $K_{\mu \nu}= \mathcal{P}^\alpha_{\; \mu} \mathcal{P}^\beta_{\; \nu} \nabla_{(\alpha} n_{\beta)}$, the round bracket ``(...)" in the subscript denotes the symmetric part of the tensor, and $\nabla_{\alpha}$ represents a covariant derivative.
This gives us
\begin{align}
[\nu^\prime]=0,\quad [\lambda]=0,
\end{align}
which are equivalent to 
\begin{align}
[p_r]=0,\quad [m]=0. \label{eq:junction_2_BC1}
\end{align}
Notice that $p_t$ is not required to be continuous.

Although we impose the above set of junction conditions (Eqs.~\eqref{eq:junction_1_BC1} and \eqref{eq:junction_2_BC1}) in this paper, one could consider a different phase separation scenario by allowing for a discontinuity in $K_{\mu \nu}$. Such a discontinuity introduces a surface stress-energy tensor, $S_{\mu\nu}$, at the interface (known as thin shells \cite{Israel_1966, Poisson_2004, Lobo_2005, Pereira_2014, Pereira_2019}):
\begin{align}
    S_{\mu\nu} = -\frac{1}{8\pi}\left([K_{\mu\nu}] - [K]h_{\mu\nu}\right).
\end{align}

For a spherically symmetric system, there are two degrees of freedom in the surface stress-energy tensor components, namely $S_{tt}$ and $S_{\theta \theta}$ ($S_{\phi \phi}$).
Here, we demonstrate a specific example, where $S_{\theta \theta}=S_{\phi \phi} = 0$ and $S_{tt}$ is chosen such that $\tilde p_+ = \tilde p_- = p_{\rm trans}$ at the interface. This introduces discontinuities in $p_r$ and $m$, similar to those in the radial perturbations of an elastic HS with a rapid phase conversion at the interface discussed in \cite{Pereira_2019}. As a result, a thin shell with finite mass appears at the interface, with the conditions: 
\begin{align}
    [\tilde p]=0,\quad [2 r \nu^\prime e^{-\lambda}]=-[e^{-\lambda}],
\end{align}
In this case, the junction conditions for $p_r$ and $m$ do not take simple closed-form expressions, but we can easily check the Newtonian limit:
\begin{align}
    [p_r]=\Pi_{r-},\quad [m^2]=-8\pi r^4[p_r],
\end{align}
where $\Pi_{r-}$ is the radial traction at the solid side of the interface (the sheared part in Eq.~\eqref{eq: pr}).
This corresponds to a thin shell with finite mass sandwiched between the two phases, where the radial pressure difference is balanced by gravity.

\section{\label{AP: karlovini}Revisiting the Elastic Effects on Maximum Mass in HSs}
In this appendix, we revisit the analysis performed by Karlovini and Samuelsson~\cite{Karlovini:2002fc} to justify the results presented in Sec.~\ref{Sec: Stellar Structure}. We resolve the structural equations for an elastic HS using the EOS defined as follows (see also Eq.~(209) in~\cite{Karlovini:2002fc}):
\begin{align}
\tilde{\rho} &= \frac{p_1}{\tilde{\Gamma} - 1} \left[\frac{\tilde{p}}{p_1} + \left(\frac{\tilde{p}}{p_1}\right)^{1/\tilde{\Gamma}}\right],
\end{align}
where $\tilde{\Gamma} = \tilde{\beta}/\tilde{p}$. Here, $\tilde{\Gamma}$ is set to 5/3, and the reference pressure $p_1$ is given as $1.5 \times 10^{37}\, \mathrm{dyn}\, \mathrm{cm}^{-2}$. The shear modulus is assumed to be proportional to the unsheared pressure $\tilde{\mu} = k\,\tilde{p}$, where $k$ is a free parameter.

\begin{figure}[!htp]
    \centering
    \includegraphics[width=8.6cm]{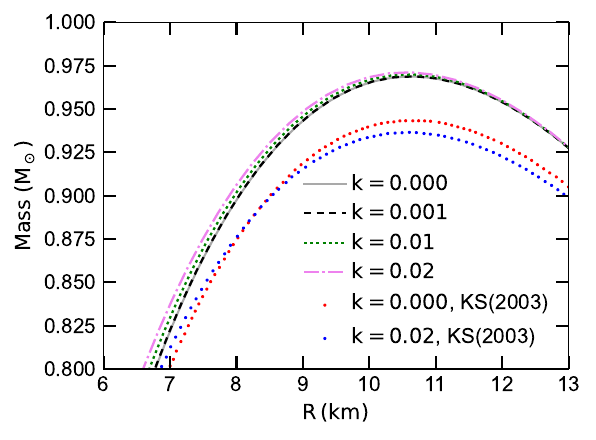}
    \caption{\label{fig: karlovini MR} $M$--$R$ relation for HSs. The curves correspond to our results with different values of $k$, while the dots represent the results from \cite{Karlovini:2002fc}.
    }
\end{figure}

Karlovini and Samuelsson~\cite{Karlovini:2002fc} reported that the maximum mass of an HS decreases with an increasing $k$. However, in Fig.~\ref{fig: karlovini MR}, we observe the opposite trend compared with their conclusions, where the maximum mass increases as $k$ increases, and the maximum mass of our results is higher than theirs. 
Our increasing trend is consistent with previous studies of anisotropic stars with $p_t > p_r$ (see, e.g., \cite{Pretel2020}).

\begin{figure}[!htp]
    \centering
    \includegraphics[width=8.6cm]{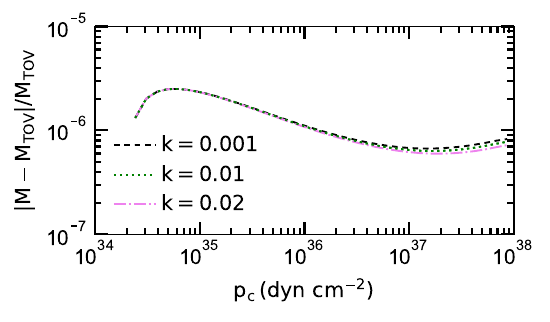}
    \caption{\label{fig: karlovini error} Fractional difference in mass between the solutions derived from Eqs.~\eqref{eq: m}--\eqref{eq: z} and the anisotropic TOV equations (Eqs.~\eqref{eq: anisotropic TOV m}--\eqref{eq: anisotropic TOV pr}), across different values of the shear modulus parameter $k$.}
\end{figure}

To further verify our results, we compared the solutions derived from the structural equations with those obtained directly from the anisotropic TOV equations (Eqs.~\eqref{eq: anisotropic TOV m}--\eqref{eq: anisotropic TOV pr}).
Starting with the $z$--profile obtained from the structural equations, which relates to the anisotropy $\sigma$ via Eq.~\eqref{eq: sigam-z}, we treated this profile as an ``effective EOS'' and solved the anisotropic TOV equations with the same boundary conditions as we mentioned in Sec.~\ref{Sec: BC}. These two sets of equations yield equivalent results. As shown in Fig.~\ref{fig: karlovini error}, the fractional difference in mass between the two sets of differential equations remains around $10^{-6}$ for various shear moduli. This minimal difference falls within an acceptable range, providing strong evidence for the reliability of our results. Furthermore, the consistency of this calculation has been verified using two independent numerical codes.

\nocite{*}

\bibliography{apssamp}

\end{document}